\documentclass[aps,prl,twocolumn,showpacs,groupedaddress]{revtex4}  % for review and submission 

\usepackage{graphicx,subfigure}  % needed for figures
\usepackage{dcolumn}   % needed for some tables
\usepackage{bm}        % for math
\usepackage{amssymb}   % for math
\usepackage{amsmath} 

\newcommand{\be}{\begin{equation}} 
\newcommand{\ee}{\end{equation}}
\newcommand{\bea}{\begin{eqnarray}}
\newcommand{\eea}{\end{eqnarray}}
\newcommand{\ud}{\mathrm{d}}
\newcommand{\intd}{\, \mathrm{d}}

\begin{document}

\title{Dynamics of gene expression under feedback}
\author{Otto Pulkkinen}
\author{Johannes Berg}
\affiliation{Institut f\"ur Theoretische Physik,
Universit\"at zu K\"oln\\
Z\"ulpicher Stra{\ss}e 77,
50937 K\"oln,
Germany}

\date{\today}

\begin{abstract} 
\noindent 
Gene expression is a stochastic process governed by the presence of
specific transcription factors. Here we study the dynamics of gene
expression in the presence of feedback, where a gene regulates its own
expression. The nonlinear coupling between input and output of gene
expression can generate a dynamics different from simple scenarios
such as the Poisson process. This is exemplified by our findings for
the time intervals over which genes are transcriptionally active and
inactive. We apply our results to the \emph{lac} system in
\emph{E. coli}, where parametric inference on experimental data
results in a broad distribution of gene activity intervals.
\end{abstract}


\maketitle
\noindent

Gene expression is a dynamic process, which transfers genetic
information from DNA to functional molecules such as
proteins~\cite{Hawkinsbook}. This process is controlled by specific
proteins, called transcription factors, which bind to DNA typically
near the starting point of a gene. Transcription factors can act as
enhancers or as repressors of gene transcription by attracting or
impeding the molecular machinery which \emph{transcribes} a gene. This
machinery, called RNA-polymerase, produces m(essenger)RNA molecules
from the DNA template. A single mRNA transcript is later
\emph{translated} to several copies of polypeptide chains, which fold
into proteins.

Due to the low copy number of the specific molecules typically present
in a cell, these processes are intrinsically stochastic. Thus genes
can be thought of as `toggling' at random points in time between
transcriptionally active and inactive states~\cite{GoldingCox:2006}.
One manifestation of this stochasticity is cell-to-cell variations of
mRNA and protein numbers in populations of genetically identical
cells~\cite{McAdamsArkin:1997}.

The time intervals over which the gene is transcriptionally active can
be very short. Frequently, only a single mRNA molecule is produced
before an enhancer molecule unbinds again from the regulatory region of a
gene, or a repressor molecule binds, causing a change in the transcriptional state of the 
gene~\cite{Yu.etal:2006}. For short gene-on times, individual mRNA molecules
are produced in statistically independent events, which can be
modelled by a Poisson process. As a result, fluctuations in mRNA numbers 
follow Poisson statistics. This picture of gene expression dynamics 
is frequently referred to as the Poisson scenario~\cite{KaufmannvanOudenaarden:2007}.

A similarly simple picture emerges if a gene is transcriptionally
active long enough to allow for multiple mRNA molecules to be
produced. At constant concentration of transcription factors, binding
and unbinding of transcription factors to DNA takes place at constant
rates. Hence the time intervals over which a gene is active or
inactive are distributed exponentially. mRNA molecules are
produced while the gene is active, leading to bursts in mRNA
numbers~\cite{KaufmannvanOudenaarden:2007}. Both exponentially distributed gene-on times, and
transcriptional bursts have recently been observed
experimentally~\cite{Golding.etal:2005,Chubb.etal:2006}.

\begin{figure}[tb!]
\includegraphics[width=0.4\textwidth]{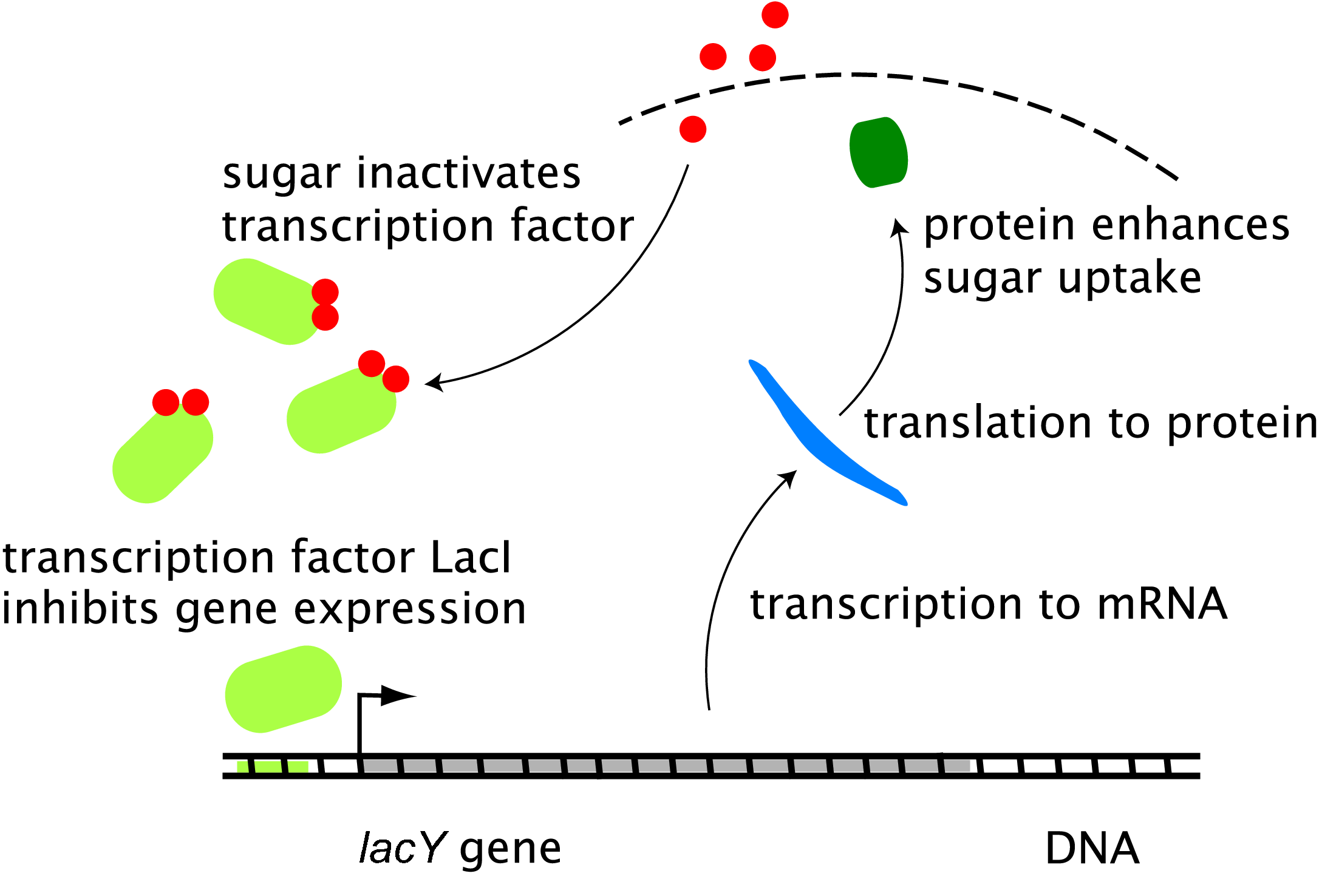}
\caption{\label{fig:lac} {\bf Feedback in the {\it lac} system.} This
  schematic picture shows transcription and translation of the lacY
  gene to a protein which facilitates the uptake of lactose (a sugar) and its chemical analogues from the
  environment. Expression of this gene is repressed by the
  transcription factor LacI.  Individual sugar molecules (\emph{e.g.}~in the form of allolactose) bind to the
  transcription factor, rendering the transcription factor less likely
  to bind to its binding site on DNA. The sugar molecules thus act as
  effective inducers of lacY expression. LacI represses also other genes ({\it lacA} and {\it lacZ}), which
  encode enzymes used to digest lactose. The feedback
  loop ensures that {\it lac} genes are repressed in the absence of
  inducing sugar molecules in the environment.}
\end{figure}
 
This simple picture of expression dynamics must break down in the
presence of feedback, which is the subject of this paper: Direct or
indirect coupling between the transcript level of a gene and its
transcription rate can introduce a nontrivial dynamics.
 
An example of feedback is direct autoregulation, which is pervasive in
bacteria. Feedback is typically non-linear, since
binding of transcription factors to DNA or to other molecules
saturates at high concentrations. Feedback can play crucial functional
roles, for instance in the \emph{lac} system, which controls the uptake of
sugar in bacteria. The gene \emph{lacY} regulates its expression by
inactivating its own repressor (a doubly-negative feedback loop, see
Fig.~\ref{fig:lac}).

The consequences of feedback for cell-to-cell variability have been
studied both experimentally~\cite{Ozbudak.etal:2004} and
theoretically~\cite{KeplerElston:2001,Hornos.etal:2005,Fournier.etal:2007}.
The analysis of the dynamics in autoregulatory systems, however, has been limited to linear
models~\cite{ViscoAllenEvans:2008}. In this Letter, we analyze the dynamics of regulatory systems with
non-linear feedback kinetics. The dynamic effects of
feedback turn out to be particularly marked in the case of the
nonlinear doubly-negative feedback, as in the \emph{lac} system, where
a broad distribution of gene-on times emerges.

We consider a system consisting of a single gene with transcriptional state $S(t)$ and number of proteins $Y(t)$ present in the cell at time $t$. The state $S=0$ stands for a transcriptionally inactive gene, and $S=1$ for an active gene. The dynamics is assumed to be Markovian. To model feedback, both the rate at which the gene goes from the inactive to the active state, $\alpha(Y)$, and from the active to the inactive state, $\beta(Y)$, can depend on the number of proteins $Y$. In the gene-on state, the gene produces mRNA molecules according to a Poisson process with rate of production $\gamma$. Each mRNA molecule is translated to a geometrically distributed number of proteins~\cite{Berg:1978}, with mean number $\rho$, before decaying. The number of proteins
$Y$ decreases at a rate $\eta$, which is largely due to dilution by cell division:
\bea
\label{rates1}
&& \qquad \qquad \quad S=0\ \substack{{\alpha(Y)} \\ \longrightarrow \\ \longleftarrow \\ \beta(Y)}\  S=1 \\
\label{rates2}
&& Y \stackrel{\gamma \cdot S}{\longrightarrow} Y + Z\ ,\quad Z \sim
\mathrm{geometric\ with\ mean}\  \rho 
\eea 
\vspace{-6mm} 
\be
\label{rates3}
\!\! Y \stackrel{\eta\cdot Y}{\longrightarrow} Y - 1 \ , 
\ee
where the rate $\gamma$ is multiplied by $S$ in the second equation to take into account that mRNA is produced in the active state only. This model assumes a separation of time scales between the mRNA and protein dynamics, with a fast production of proteins from mRNA, and neglects any time delays between a change in the protein level and its effect on the gene activation and inactivation rates. Models that explicitly include the number of mRNA molecules lead to very similar results as tested in numerical simulations.

\begin{figure}[tbh]
\includegraphics[width=0.44\textwidth]{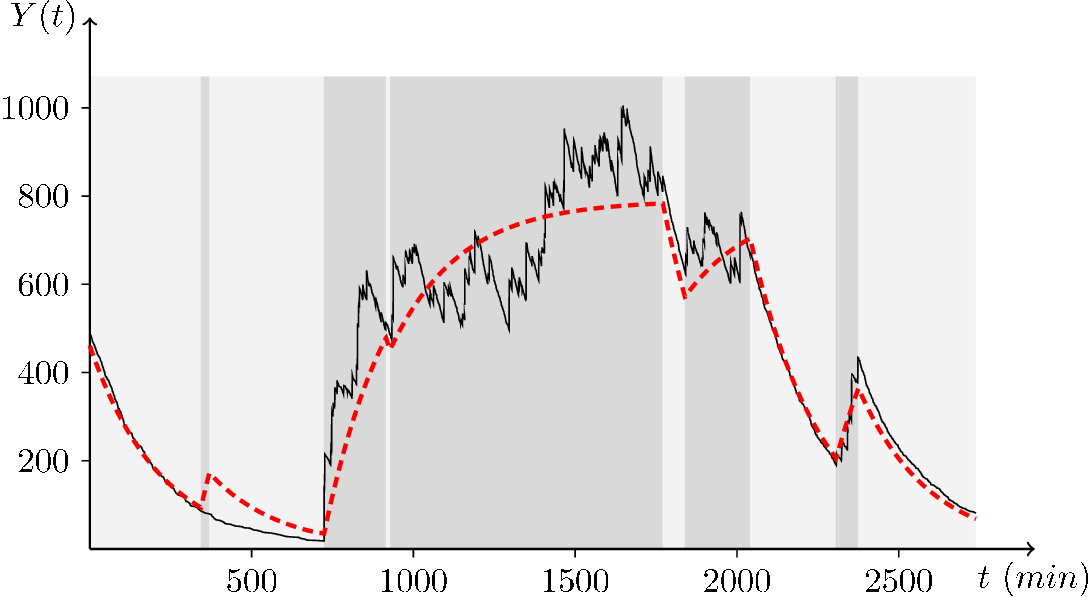}
\caption{\label{fig:sample}{\bf Sample path of the model.}  The solid
  curve shows the number of proteins $Y$ as a function of time. The
  number of proteins tends to increase when the gene is active ($S=1$,
  dark background), and decreases when the gene is inactive ($S=0$,
  light background). The dashed curve shows the piecewise
  deterministic approximation, see text. The sample path includes an
  instance where the gene is on for a brief time at ca. 330 minutes,
  but no transcription takes place.  The model parameters were
  inferred for the {\it lac} system, see footnote \cite{endnote22}. 
  The lengths of the gene-on intervals can exceed the cell division 
  time (here 216 min) because the protein concentrations are 
  inherited~\cite{Golding.etal:2005}.  }
\end{figure}

Fig.~\ref{fig:sample} shows a sample path of the model indicating both
the number of proteins $Y$ and the transcriptional state of the
cell. Fig.~\ref{fig:sample} also shows the result of a \emph{piecewise
  deterministic approximation}, which assumes a continuous
deterministic increase of the number of proteins at rate
$\gamma \rho - \eta Y(t)$ when the gene is on, and
an analogous decrease at rate $\eta Y(t)$ when the gene is off.

The piecewise deterministic approximation suggests a self-contained
model, which retains the transitions between the transcriptional
states as the only source of stochasticity. The protein level dynamics
$Y(t)$ in this approximative model thus consists of deterministic
exponentially increasing and decreasing paths joined together at
randomly positioned switching times of the transcriptional state
$S$. The switching times themselves depend on the values of $Y(t)$
because of the feedback. Such models are known as \emph{piecewise
  deterministic Markov processes} in the mathematical
literature~\cite{Boxma.etal:2005} and, specifically to describe
feedback, \emph{feedback fluid
  queues}~\cite{ScheinhardtvanForeestMandjes:2005}. In the context of
gene regulation, such approximations have been considered by Kepler
and Elston \cite{KeplerElston:2001}.

The forward Kolmogorov (master) equation of the piecewise deterministic approximation reads 
\bea
\label{pwdet_fwd_0}
\partial_t p_0 (y,t) &=& -\partial_y \lbrack (-\eta y) p_0 (y,t) \rbrack  \\
& & \qquad+ \beta(y) p_1(y,t) - \alpha(y) p_0(y,t) \nonumber \\
\label{pwdet_fwd_1}
\partial_t p_1 (y,t) &=& -\partial_y \lbrack (\gamma \rho -\eta y) p_1 (y,t) \rbrack  \\
& & \qquad +\alpha(y) p_0(y,t) - \beta(y) p_1(y,t) \ , \nonumber 
\eea
where $p_i (y,t)$ denotes the probability of the event $\lbrace
S(t)=i, Y(t)=y \rbrace$. In this approximation, $y$ has an upper bound
$\Delta := \gamma \rho/\eta$, which is determined by the balance of protein
production and degradation in the gene-on state. Under the
biologically reasonable assumption that the rate functions $\alpha$
and $\beta$ are bounded away from zero on $[0,\Delta]$, the stationary
solution to Eqs.~(\ref{pwdet_fwd_0}) and (\ref{pwdet_fwd_1}) reads 
\be
\label{pwdet_stat_0}
\phi_0 (y) = \frac{1}{Z} \, y^{ -1}  \exp \bigg\lbrack - \int \left( \frac{\beta(y)}{\gamma \rho -\eta y} -\frac{\alpha(y)}{\eta y}\right) \intd y \bigg\rbrack \\
\ee
\be
\label{pwdet_stat_1} 
(\Delta - y) \phi_1 (y) = y\phi_0 (y) \ ,
\ee 
where the subindices again
refer to the transcriptional state. The marginal distribution for the number of proteins $Y$ is given by the sum of the distributions in Eqs.(\ref{pwdet_stat_0}) and (\ref{pwdet_stat_1});
 \be
 \label{f_def}
 f(y) := \phi_0(y) + \phi_1 (y) \ .
 \ee
The constant $Z$ in Eq.~(\ref{pwdet_stat_0}) normalizes this
distribution.

In the following, we derive the distribution of gene-on times in a
stationary process, the analysis of gene-off times being analogous. We
define $T$ as the moment of the first gene inactivation after time
$t_0=0$ in a process that was started at $t=-\infty$, and consider
only those paths of the process that have an activation event taking
place immediately after $t_0$. The probabilities conditional on the 
occurrence of such an event are known in the 
mathematical literature as Palm probabilities \cite{BaccelliBremaudbook} 
and denoted by $P^0$. For example, one obtains for the probability that 
the protein level at the time of activation is less than or equal to $y$ that \cite{Boxma.etal:2005}
\be 
\label{P0_Y}
P^0(Y(0)
\leq y) = Z_0^{-1} \int_0^y \alpha(y)\phi_0(y)\intd y \ , 
\ee 
where $Z_0 = \int_0^{\Delta} \alpha(y) \phi_0(y)\, \mathrm{d} y$ is a normalization constant.
The probability that the gene is active longer than for a given time $\tau$ is then 
\bea
\label{FbarT}
&& \!\!\!\!\!\! P^0( T > \tau ) = E^0 \exp\left( - \int_0^{\tau} \beta( \tilde{Y}(t) )
  \intd t \right) \\
\label{FbarT_alt}
&& \!\!\!\!\!\! = Z_0^{-1}\!\! \int_0^{\Delta} \!\! \alpha(y)\phi_0(y) \exp \Big\lbrack - \int_y^{\Delta-(\Delta-y)\mathrm{e}^{-\eta\tau}} \!\!\!\!\!\!\! \frac{\beta(z)}{\gamma \rho - \eta z} \intd z \Big\rbrack  \intd y. \nonumber\\
\eea 
In Eq.~(\ref{FbarT}), $\tilde{Y}(t) = \Delta - (\Delta -
Y(0))\mathrm{e}^{-\eta t}$ is an exponentially increasing trajectory with a
random initial state $Y(0)$, and $E^0$ is the expectation with respect
to $P^0$. Eq.~(\ref{FbarT_alt}) follows from Eq.~(\ref{FbarT}) by a change of variable.

We evaluate these quantities for a concrete example based on the {\it
  lac} system. The activation rate $\alpha$ is taken to be independent
of $Y$ as a first approximation (allowing for $Y$-dependent activation
rates~\cite{ElfLiXie:2007} does not change the general conclusions),
whereas the gene inactivation rate $\beta$ depends of protein level
$Y$: At high protein levels, the cell takes up sugar molecules from
the environment at a high rate, leading to a high steady state
concentration of sugar in the cell~\cite{Ozbudak.etal:2004}. This
leads to a decreased number of active repressors, as discussed in
Fig~\ref{fig:lac}.  The gene inactivation rate can be written as the
product of the binding rate $b$ for active repressors and the fraction
of active repressors $\lambda(Y)$, which is of the Michaelis-Menten
form \cite{Mettetal.etal:2006}
\be
\label{LacMM}
\beta(Y) = b\lambda(Y) = \frac{b}{1+(A+BY)^2} \ .  
\ee 
Here $A$ and $B$ describe the passive and LacY-dependent active uptake of the inducing sugar from the environment respectively~\cite{Mettetal.etal:2006}. The second power in
Eq.~(\ref{LacMM}) arises because two inducers bound to a repressor are
needed to prevent the repressor from binding to DNA
\cite{YagilYagil:1971}. The inactivation rate~(\ref{LacMM}) thus
results in a nonlinear regulatory feedback.

\begin{figure}[tbh]
\includegraphics[width=.4\textwidth]{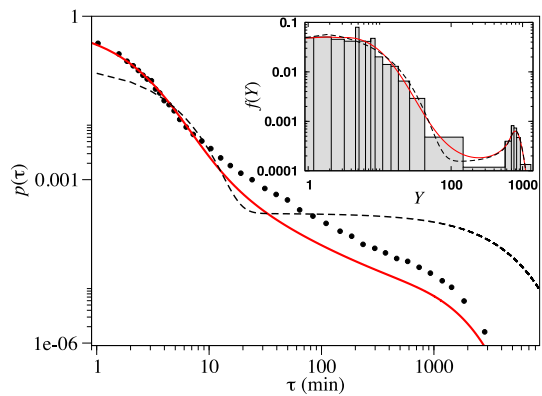}
\caption{\label{fig:otd} {\bf Distribution of gene on-times in the
    \emph{lac} system.}  The solid curve gives the probability density
  function, calculated from Eq.~(\ref{FbarT_alt}), for the lengths of
  time intervals over which the gene is on under the piecewise
  deterministic assumption. A similarly broad distribution is found in
  numerical simulations of the model~(\ref{rates1})--(\ref{rates3})
  (black dots). For comparison, the dashed line gives a mixture of two
  exponential distributions corresponding to a mixture of induced and
  uninduced cells without feedback. Inset: The histogram shows the
  distribution of lacY levels measured by the van~Oudenaarden
  lab~\cite{Ozbudak.etal:2004}. Bin widths were chosen to contain
  equal number of data points. The solid curve gives the corresponding
  analytical result of Eq.~(\ref{f_lac}), and the dashed curve the
  result (\ref{f_mix}) for the exponential model without feedback, with
  parameters fit to the data in both cases.}
\end{figure}

Fig.~\ref{fig:otd} shows the distribution $p(\tau)= -\ud P^0 (T>\tau)/
\ud \tau$ of gene-on times in the \emph{lac} example. The model parameters were
inferred from experimental data of the van~Oudenaarden lab~\cite{Ozbudak.etal:2004} 
as explained in the footnote~\footnote{
  The experimental data used in the inset of Fig.~\ref{fig:otd} gives the
  lacY distribution at a concentration of $15\ {\mu}\mathrm{M}$ of the
  inducing sugar TMG. The fluorescence data of~Fig.~2b in~\cite{Ozbudak.etal:2004}
  was rescaled to a steady state protein level of $Y=790$ proteins per
  cell at full induction as determined in~\cite{Ozbudak.etal:2004}.  The
  parameter inference was carried out by convoluting the stationary
  solutions (\ref{f_mix}) and (\ref{f_lac}) with a Gaussian distribution of mean zero and a
  standard deviation which depends linearly on $Y$.  This accounts for
  the inevitable smearing out of the divergences of the stationary protein level distributions by
  fluctuations of mRNA and protein numbers neglected in the piecewise deterministic approximation, as well as by
  experimental noise. The inferred parameters of~(\ref{rates1})--(\ref{rates3}) and
  (\ref{LacMM}) for the piecewise deterministic model with feedback are $\alpha/\eta = 1.59$, $b/\eta = 192$, and
  $B=0.035$. The value of $A$ was found to be negligible. For the simple mixture of cells, the inferred values are $\alpha/\eta = 0.9$, $b/\eta = 80$, $b\lambda(\Delta) = 0.08$, and $r=0.35$. The
  numerical simulations in Figs.~\ref{fig:sample} and~\ref{fig:otd}
  used $\eta^{-1}$ determined by the cell cycle time of 216 min given
  by~\cite{Mettetal.etal:2006} for the particular strain of {\it
    E.~coli} used in~\cite{Ozbudak.etal:2004,Mettetal.etal:2006}. The
  average number of proteins produced from one mRNA molecule was taken
  to be 35, as estimated in~\cite{Mettetal.etal:2006}.}. 
Both the result of the piecewise deterministic approximation (solid
line), and numeric simulations of the full model
(\ref{rates1})--(\ref{rates3}) (black dots) exhibit a broad distribution of 
gene-on intervals, with an exponential cut-off due to saturating inducer concentrations. 
The small difference between the gene-on distributions between the two
models stems from enhancing upward fluctuations at high protein
numbers, which are absent in the piecewise deterministic approximation.

These results can be contrasted with a simple model lacking feedback, where 
gene-on times are exponentially distributed. We consider a mixed 
population of cells with a fraction $r$ of cells having gene 
inactivation rate $\beta(0)$, and a fraction $1-r$ having rate $\beta(\Delta)$. The dashed line 
in Fig.~\ref{fig:otd} shows the corresponding gene-on time distribution $r
\beta(0)e^{-\beta (0) \tau}+ (1-r)\beta(\Delta)e^{-\beta(\Delta)\tau}$. 
The stationary protein level distributions of the simple mixture of cells reads
\be
\label{f_mix}
f_{\mathrm{mix}} (y) = r f_{\beta(0)}( y ) + (1-r)f_{\beta(\Delta)}(y)
\ee
with
\be
 f_{c}( y ) = B(\alpha/\eta,c/\eta ) ^{-1}
\Delta^{1-\frac{\alpha+c}{\eta}}y^{\frac{\alpha}{\eta} -1} \left( \Delta  - y \right)^{\frac{c}{\eta} -1},
\ee
where $B$ denotes the Euler Beta function. This and the corresponding distribution (\ref{f_def}) for the piecewise deterministic approximation,
\be
\label{f_lac}
f (y) = \frac{ \lambda(y)^{\frac{\kappa}{2}}}{Z}
y^{\frac{\alpha}{\eta} -1} \left( \Delta - y \right)^{\kappa-1} {\mathrm e}^{-\kappa(A+B\Delta) \arctan (A+By)},  
\ee
where $\kappa = \beta(\Delta)/\eta$, can both be accurately fitted to the histogram of LacY levels
experimentally measured in a population of \emph{E. coli}
cells~\cite{Ozbudak.etal:2004}, see the inset of 
Fig.~\ref{fig:otd}. However, the \textit{dynamics} of the
transcriptional state is markedly different in the two models, as is 
evident from the gene-on times in Fig.~\ref{fig:otd}.
This shows how dynamic information, now within experimental reach~\cite{Yu.etal:2006}, 
can distinguish between systems with similar statistics of gene expression levels.     

In summary, we have shown how feedback shapes the dynamics of
regulatory networks. This dynamics also characterizes transitions in
multistable regulatory systems.  The hysteretic transition from an
uninduced state to an induced state of the
\emph{lac}-system~\cite{Ozbudak.etal:2004} is accompanied by a
divergence of time intervals over which the lacY-gene is
transcriptionally active: At low concentrations of the inducing sugar,
lacY is active only for short periods of time governed by the rate of
repressor binding. At increasing concentrations, the gene on-time
distribution broadens (Fig.~\ref{fig:otd} and Eq.~(\ref{FbarT_alt})),
and finally at high concentrations of the inducing sugar the gene is
transcriptionally active most of the time.  Viewing the
transcriptional state $S(t)$ of the gene as a continuous field of
discrete spin variables in 1D, this behaviour can be seen as a
divergence of magnetic domain sizes: Feedback introduces interactions
between spins at different times, with a range determined by the
protein life-time $\eta^{-1}$, which also determines the eventual
exponential cut-off in the distribution of gene-on times.

We have focused on a doubly-negative feedback-loop based on
the \emph{lac}-system.  Positive feedback also generates non-trivial
dynamics, however, it is the inverse of gene-off times which turn out
to follow a broad distribution. Our analysis is not restricted to
systems with direct autoregulation. Regulatory networks typically
contain loops generating correlations between different events
affecting the transcription of a gene. These correlations are at the
heart of deviations from statistical pictures such as the Poisson
scenario.  An example is feed-forward loops, where the signal from one
gene is recombined with a time-delayed copy of itself to produce a
simple filter~\cite{Shen-Orr.etal:2002}. Feedback loops also play a
key role in the control of ionic channels determining the excitability
of the heart, where non-exponential distributions of channel-open and
-closed times have been observed~\cite{Colecraft.etal:2002}.

\begin{acknowledgments}
Many thanks to Alexander van Oudenaarden for the experimental data
in Fig.~\ref{fig:otd}, and to Ulrich Gerland and Georg Fritz for discussions.
Funding from the DFG is acknowledged under
grant BE 2478/2-1 and SFB 680.
\end{acknowledgments}

%\bibliography{./regulatory}

\end{document}